# Combining smart card data and household travel survey to analyze jobs-housing relationships in Beijing

Ying LONG[a,*], Jean-Claude THILL[b]

[a]. Beijing Institute of City Planning, Beijing 100045, China

[b] Department of Geography and Earth Sciences, The University of North Carolina at Charlotte, Charlotte NC 28223, USA

[*] Corresponding author at: Beijing Institute of City Planning, Beijing 100045, China. Tel.: +86 10 88073660. Fax: +86 10 68031173

E-mail address: longying1980@gmail.com (Y. Long)

Abstract: Location Based Services (LBS) provide a new perspective for spatiotemporally analyzing dynamic urban systems. Research has investigated urban dynamics using GSM (Global System for Mobile Communications), GPS (Global Positioning System), SNS (Social Networking Services) and Wi-Fi techniques. However, less attention has been paid to the analysis of urban structure (especially commuting pattern) using smart card data (SCD), which are widely available in most cities. Additionally, ubiquitous LBS data, although providing rich spatial and temporal information, lacks rich information on the social dimension, which limits its in-depth application. To bridge this gap, this paper combines bus SCD for a one-week period with a one-day household travel survey, as well as a parcel-level land use map to identify job-housing locations and commuting trip routes in Beijing. Two data forms (TRIP and PTD) are proposed, with PTD used for jobs-housing identification and TRIP used for commuting trip route identification. The results of the identification are aggregated in the bus stop and traffic analysis zone (TAZ) scales, respectively. Particularly, commuting trips from three typical residential communities to six main business zones are mapped and compared to analyze commuting patterns in Beijing. The identified commuting trips are validated on three levels by comparison with those from the survey in terms of commuting time and distance, and the positive validation results prove the applicability of our approach. Our experiment, as a first step toward enriching LBS data using conventional survey and urban GIS data, can obtain solid identification results based on rules extracted from existing surveys or censuses.

Keywords: bus smart card data; jobs-housing relationships; commuting trip; household travel survey; Beijing

# 1 Introduction

This paper identifies jobs-housing locations and commuting patterns in Beijing using smart card data (SCD), which stores the daily trip information of bus passengers. Related researches analyzing jobs-housing location have used data acquired through questionnaires, surveys or censuses. The increasing pervasiveness of location based services (LBS) associated with the prevalence of location-acquisition technologies, including GSM (Global System for Mobile Communications), GPS (Global Positioning System), SNS (Social Networking Services), as well as Wi-Fi (wireless fidelity), has led to the creation of large-scale and high-quality spatiotemporal datasets. This development has also created opportunities to better describe and understand urban structures in multiple dimensions. These datasets are profoundly significant for analyzing urban and environmental systems, e.g. jobs-housing relationships (Batty, 1990). Meanwhile, geo-tagged smart card is an effective alternative tool for individual data acquisition necessary to analyze urban spatial structures.

Urban and environmental studies using fine-scale individual data have been the focus in this area of research. Goodchild (2007) first proposed the concept of "humans as sensors", in contrast to traditional optical remote sensors, to acquire spatial and social information. Volunteered data have been extensively applied to analyze urban structures in conjunction with GPS, A-GPS, GSM, SNS and Wi-Fi technologies (Ahas and Mark, 2005). With respect to GPS, Newhaus (2009) used location data to record and visualize urban diaries. Jiang (2009) analyzed taxi passengers' travel characteristics and identified the influence of urban structure on those characteristics using GPS-recorded taxi trajectories in Sweden. Liu et al (2010) identified taxi drivers' behavior patterns from their daily digital records. With respect to GSM network (see Steenbruggen et al, 2011 for a review), Ratti et al (2006) evaluated the density and spatiotemporal characteristics of urban activities using mobile phone data in Milan, Italy. Calabrese and Ratti (2006) monitored urban activities using mobile phone data in Rome, Italy. Reades (2009) visualized urban activities using mobile phone data. With respect to SNS, Milton (2010) mapped snow depth in the UK using Twitter data. Jones et al (2010) created a blog on theatre design in New York, and found that blog data can be applied to effectively analyze economic-geographic phenomena. As for Wi-Fi, Rekimoto et al (2007) used Wi-Fi-based location detection technology to log the locations of device holders from received Wi-Fi beacon signals, a technology that works both indoors and outdoors. Torrens (2008) developed a system to detect Wi-Fi infrastructure and transmission and analyze their geographic properties, and tested this system in Salt Lake City, Utah, USA. Meanwhile, the discipline of time geography established by Hagerstrand (1970) also benefited from the development of LBS by retrieving more objective data. In sum, various LBS technologies have been successfully applied in urban studies in response to the proposal by Shavol (2007) regarding human sensing. However, these technologies remain immature, and most urban structure related research employs data from urban physical space or questionnaire surveys. The acquisition of large-scale micro data remains limited (Long et al, 2010).

The smart card recording full cardholder bus trip information is an alternative form of location-acquisition technology. Smart card automated fare collection systems are increasingly applied in public transit systems. Simultaneously with collecting revenue, such systems can produce intensive travel patterns of cardholders, data that are useful for analyzing urban dynamics. Since 1990 the use of smart cards has become significant owing to the development of the Internet and the increased complexity of mobile communication technologies (Blythe, 2004). Intelligent Transportation Systems (ITS) that incorporate smart card automated fare systems either existed or were being established in over 100 Chinese cities as of 2007 (Zhou, 2007). The data generated by smart card systems track the detailed onboard transactions of each cardholder. We argue that smart card technology can provide valuable information because it is a continuous data collection technique that provides a complete and real-time bus travel diary for all bus travelers. SCD can be used to validate traditional travel models applied to public transit. In contrast to smart card data collection, conventional household travel surveys have the drawbacks of being expensive and infrequent. Notably, transit SCD collects data in fundamentally the same way as the AVI (automatic vehicle identification) system, which has been widely used in the USA to automatically identify vehicles. AVI is used in some states in the US for planning purposes. One example is New York, where the EasyPass tag is used as part of the TransMit system.

Previous studies advocate using SCD to make decisions on the planning and design of public transportation systems (see Pelletier et al (2011) for a review). In South Korea, Joh and Hwang (2010) analyzed cardholder trip trajectories using bus smart card data from ten million trips by four million individuals, and correlated this data with land use characteristics in the Seoul Metropolitan Area. Jang (2010) estimated travel time and transfer information using data on more than 100 million trips taken in Seoul on the same system. Roth et al (2011) used a real-time "Oyster" card database of individual traveler movements in the London subway to reveal the polycentric urban structure of London. Gong et al (2012) explored spatiotemporal characteristics of intra-city trips using metro SCD on 5 million trips in Shenzhen, China. However, less attention has been paid to using SCD to analyze jobs-housing relationships and commuting patterns in a metropolitan city, which is the main focus of this paper.

This paper regards jobs-housing relationships and commuting pattern analysis as a showcase for applying SCD to urban spatial analysis. The jobs-housing locations, their relationships, and working trips can be identified from SCD using Beijing as a case study. The identification results are validated using household travel survey data from Beijing. Our contention is that SCD is a good substitute for household travel surveys, or at least a complement. This paper is organized as follows. The jobs-housing trips in conventional household travel surveys are discussed in Section 2, and the SCD and other related datasets used in our research are illustrated in Section 3. The approaches for identifying housing and jobs locations, as well as commuting trips are elaborated in Section 4. In Section 5, the results of jobs-housing identification and commuting patterns are shown and analyzed in detail. Finally, we discuss our work and present concluding remarks in sections 6 and 7, respectively.

**2 Jobs-housing trips in conventional household travel surveys**

Household travel survey has been a main means of data-collection for acquiring information on resident travel behavior and exploring urban transit systems (Beijing Transportation Research Centre, 2009). The survey tracks traveler socio-economic attributes, as well as trip origin and destination, time and duration, and purpose and mode. On the one hand, traveler housing and job location are directly recorded in the survey together with his/her socio-economic attributes, and both locations are mostly aggregated in the traffic analysis zone (TAZ) scale since geographic information systems (GIS) are still rarely used to track the precise locations of individual households during surveys. On the other hand, trips between work and home (i.e. commuting trips) can be screened using the purpose attribute. These trips are also recorded using the inter-TAZ scale rather than a fine spatial scale. Therefore, jobs-housing locations and trips have already been recorded in conventional household travel surveys, but mostly using the TAZ scale. Additionally, only a small portion of all households in a given city were surveyed due to time and cost constraints.

Compared with household travel surveys, mining the enormous volume of SCD can provide a more precise spatial resolution and a larger sample, despite the SCD being unable to directly provide jobs-housing locations and commuting trips. We will focus on using SCD to identify jobs-housing relationships. Commuting patterns of trips from typical residential communities or to typical business zones can be visualized by identifying the results on the fine scale, which is not available in travel surveys in Beijing because a residential community or business zone is generally smaller than one TAZ. A more detailed commuting pattern is expected to reveal fresh information on jobs-housing relationships in a megacity. A shortcoming of SCD is that it includes no information on cardholder socioeconomic attributes, and the purpose of individual trips is also unknown. Conventional household travel surveys can supply such additional information for use in analyzing SCD, and combining SCD with household travel survey data is a promising method of jobs-housing analysis, which will be elaborated below.

**3 Data**

*3.1 Bus routes, bus stops, and traffic analysis zones (TAZs) of Beijing*

GIS layers of bus routes and stops are essential for geocoding and mapping SCD. There are 1,287 bus routes (Figure 1a) in the Beijing Metropolitan Area (BMA), which totals 16,410 km$^2$. These 1,287 bus routes have 8,691 stops (see Figure 1b). Note that a pair of bus platforms on opposite sides of a street is considered a single bus stop. For instance, there are two bus platforms at Tian'anmen Square, one on the south side of Chang'an Avenue and one on the north side. In the bus stop GIS layer, the two platforms are merged into a single bus stop feature. The average distance between each bus stop and its nearest neighbor is 231m.

We use Beijing TAZs data to aggregate the analytical results for better visualization. Totally 1,118 TAZs are defined (see Figure 1c) according to the administrative boundaries, main roads, and the planning layout in the BMA.

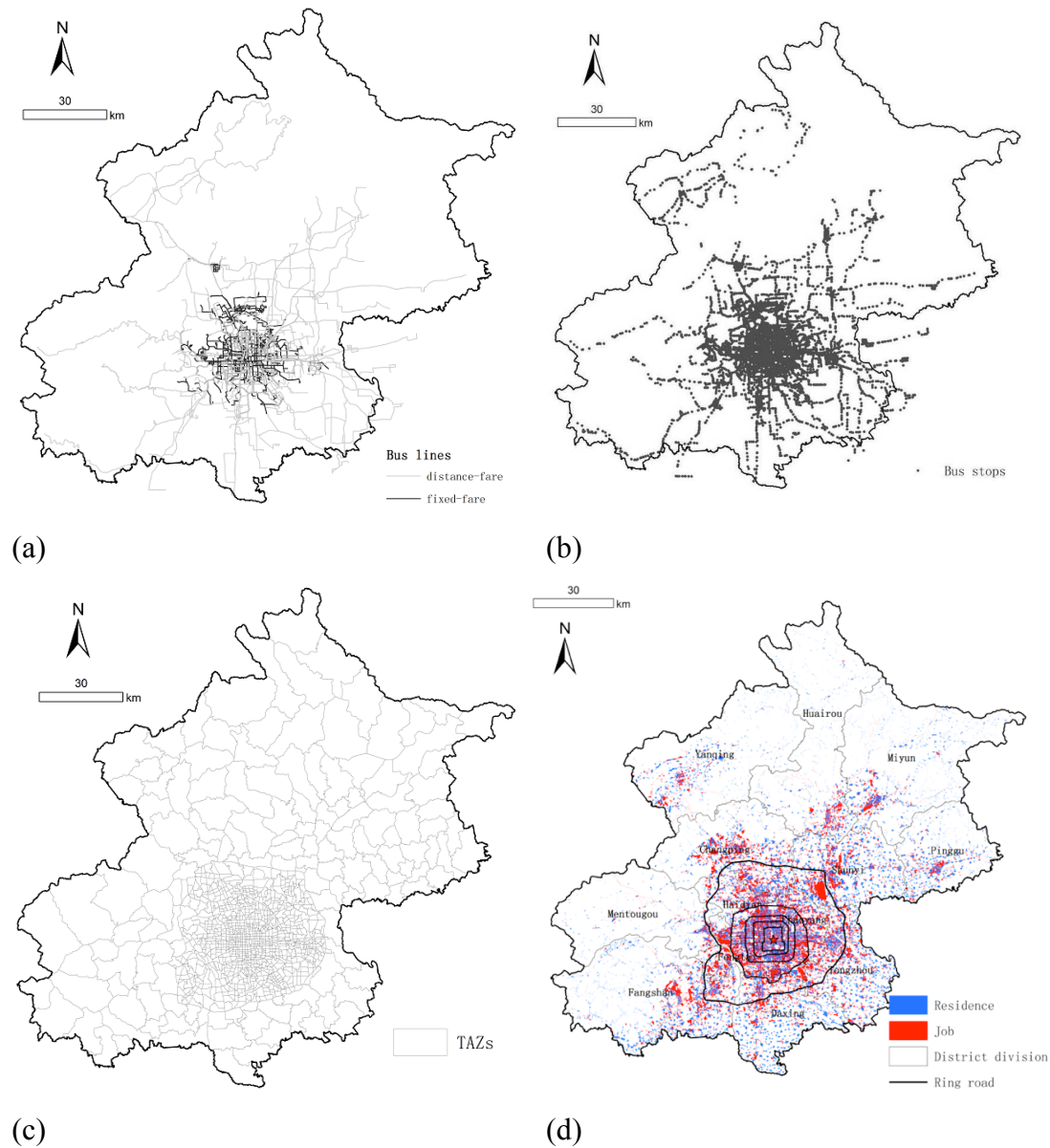

Figure 1 Bus routes (a), bus stops (b), traffic analysis zones (TAZs) (c), and land use patterns (d) of the BMA

Note: All four maps are from the Beijing Institute of City Planning. Some bus routes and stops are outside the BMA, as shown in (a) and (b), since some residents live outside the BMA and in adjacent towns in Hebei province. In the TAZ-level analysis conducted in this paper, trips outside the BMA in the SCD are not counted since there are no TAZs outside the BMA. The five circles in (d) represent the second, third, fourth, fifth, and sixth ring roads of Beijing. The star in the central area represents Tian'anmen Square.

### *3.2 Land use pattern of Beijing*

The parcel-level land use pattern data and land use type attribute of the BMA is introduced into our research to identify housing and job locations. Floor area data are available for each parcel. We assume that the residential land-use type represents housing locations, and that the commercial, public facility and industrial land-use type represents job locations. The 133,503 parcels include 29,112 residential parcels, and 57,285 parcels with job locations (labeled "job parcels" in this paper) (Beijing Institute of City Planning, 2010). Land use pattern is used to calculate the probability of each bus stop servicing a housing or job location.

### *3.3 The one-week smart card data (SCD) from Beijing in 2008*

A smart card system has been applied in the public transit system of Beijing since April 1, 2006 (Liu, 2006). The system can automatically track cardholder bus trip information. The bus share of total trips taken in Beijing during 2008 was 28.8%, and the subway share was 8.0% (Beijing Transportation Research Center, 2009). Over 10 million smart cards (see Figure 2) have been issued in Beijing since April 2, 2007, and over 90% of all bus passengers are cardholders[1]. The smart cards used by the Beijing public transit system are anonymous and users provide no personal information when applying for them. Consequently, using smart card data for research purposes involves no privacy issues.

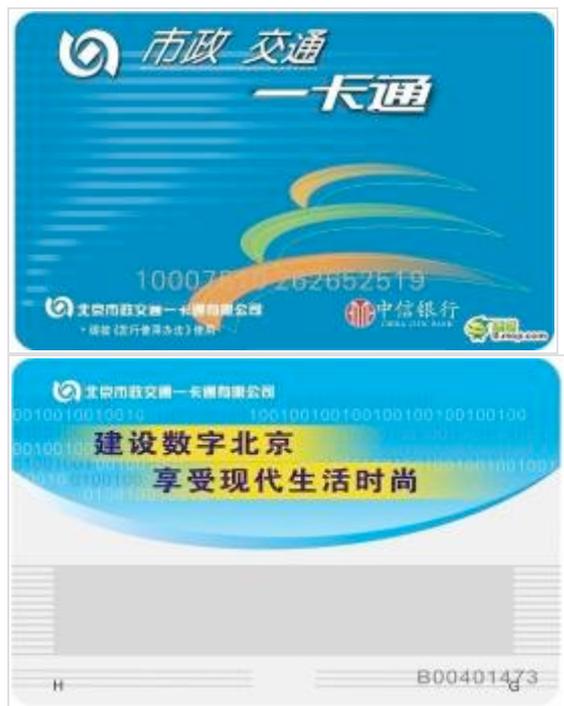

(a)                                    (b)

Figure 2 The bus smart card of Beijing (a: front side; b: back side)

---

[1] The bus smart card in Beijing is operated by Beijing Municipal Administration & Communications Card Co., Ltd. The official website is www.bjsuperpass.com. Cardholder bus trips over several weeks can be queried via this website by inputting the card ID from the front side of the card (Figure 2).

The SCD we obtained in Beijing covered a one week period in 2008 (April 7-13)[2], and did not include subway records. The SCD records essential fields (see Table 1). The data comprise 77,976,010 bus trips by 8,549,072 cardholders, and thus each cardholder makes an average of 1.30 bus trips per day.

Table 1 Data structure of the SCD

| Variable | Exemplified Values |
| --- | --- |
| Card ID | "10007510038259911", "10007510150830716" |
| Card Type | 1, 2, 3, 4 |
| Trip ID | 25, 425, 9 |
| Route ID | 602, 40, 102 |
| Route Type | 0, 1 |
| Driver ID | 11032, 332 |
| Vehicle ID | 111223, 89763 |
| Departure Data | 2008-04-08 |
| Departure Time | "06-22-30", "11-12-09" |
| Departure Stop | 11, 5, 14 |
| Arrival Time | "09-52-05", "19-07-20" |
| Arrival Stop | 3, 14, 9 |

Note: 0 stands for a fixed-fare route and 1 stands for a distance-fare route for the attribute "Route Type". For the attribute "Card Type", 1, 2, 3, and 4 denote normal, student, staff and month pass, respectively. The attribute "Trip ID" represents the accumulated trip count of a card since its issue, including both subway and bus journeys.

Two bus fare types exist, the first is fixed-fare, which is associated with short routes, and the other is distance-fare, which is associated with long routes (see Table 2 for the comparison). For the first type, 0.4 CNY is charged for each single bus trip, and the corresponding SCD contains only the departure time and stop ID and no arrival time and stop ID. Cardholders' spatiotemporal information is incomplete for this kind of route. For the latter type, the fare depends on the route ID and trip distance, and the SCD contains full information. We consider the SCD of both fare types, including fixed-fare and distance-fare. The fixed-fare trips with partial travel information lead to failure to identify a housing or job location. For instance, if the first trip of a cardholder on a given day is a fixed-fare trip, it is impossible to identify his/her housing location since the location where they first boarded is unknown. However, in some cases, such as when a fixed-fare trip is taken as a transfer between two distance-fare trips, it is still possible to identify housing location.

---

[2] In April 2008, the policy of weekly 'no driving days' for car owners had not yet been implemented.

Table 2 Comparisons between fixed-fare and distance-fare routes and SCD

|  | **Fixed-fare** | **Distance-fare** |
| --- | --- | --- |
| Route count | 566 | 721 |
| Total length (km) | 7,529.1 | 25,812.6 |
| Average length (km) | 13.3 | 35.8 |
| Trip count | 50,916,739 (65.3%) | 27,059,271 (34.7%) |

The temporal and spatial dimensions of the SCD are illustrated here. With respect to the temporal dimension of all trips in the SCD, the total count of bus trips for each day in a week is plotted in Figure 3a. The total number of bus trips on a weekday (Tuesday) and a weekend day (Saturday) for each departure hour is listed in Figure 3b, and shows significant differences between the two days. Most bus trips are distributed from 6:00 to 22:00 and match peak hours in the 2005 survey. With respect to the spatial dimension, the total daily bus trip density in the inner area exceeds that in the outer area in terms of boarding location (Figure 3c).

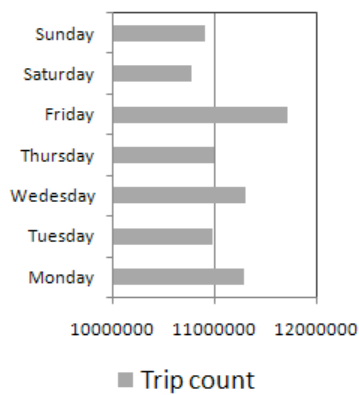
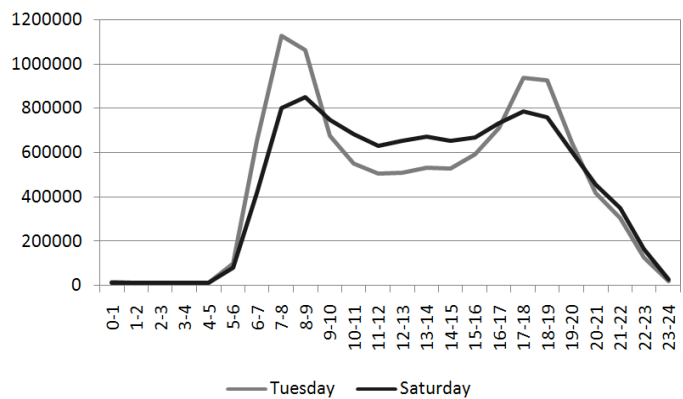

(a)   (b)

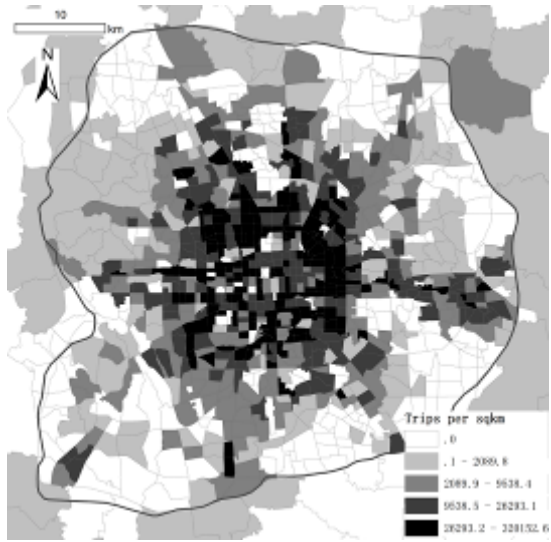

(c)

Figure 3 Spatiotemporal characteristics of bus trips in the SCD. (a) The total count of bus trips on each day of the one week study period; (b) The trip count for various departure hours on Tuesday and Saturday; (c) Trip count densities for each TAZ within the 6$^{th}$ ring road of the BMA.

*3.4 The Beijing household travel survey*

Household travel surveys for Beijing have been conducted in 1986, 2000 and 2005. The 2005 Beijing household travel survey (hereafter called the 2005 survey) is included in this paper to set rules for identifying jobs-housing locations and commuting trips. This survey covers the whole BMA, including all 18 districts with 1,118 TAZs (these TAZs are the same as the TAZs in Fig. 3) (Beijing Municipal Commission of Transport and Beijing Transportation Research Center, 2007). The sampling size is 81,760 households/208,290 persons, with a 1.36% sampling rate. This survey adopts a travel diary form. For each trip, the survey records the departure time/location, arrival time/location, trip purpose and mode, as well as other important information like trip distance, destination building type, and transit route number.

The survey also includes household and personal information. The household information includes household size, Hukou (official residence registration) status and residence location, while the personal information includes gender, age, household role, job type and location, and whether the respondent has a driving license or transit monthly pass. Additionally, the trip purposes include: 1) work, 2) school, 3) returning home from school or work, 4) Returning-home trip, 5) shopping, 6) entertainment, 7) daily life (such as dining, medical, social visit, leisure/fitness, and pick up/ delivery), 8) business, 9) other. Job types include: 1) worker, 2) researcher, 3) office/public employee, 4) teacher, 5) student, 6) self-employed, 7) household attendant, 8) retiree, 9) specialized worker (such as medical staff, professional driver, and soldier/police), 10) farmer, 11) unemployed, 12) other. Trip modes include: 1) walk, 2) bicycle, 3) electric bicycle, 4) motorbike, 5) bus, 6) mini bus, 7) metro, 8) employer-provided bus, 9) private car, 10) employer-provided car, 11) legal and illegal taxi. Among all these transportation modes, bus accounts for 13.81% of all trips in the BMA according to this survey.

**4 Approach**

*4.1 Data pre-processing and data forms*

The SCD, as the raw data recording cardholders' bus riding information, need to be pre-processed to facilitate the jobs-housing analysis and evaluate bus trips spatiotemporally. First, we geocode the SCD by linking the bus stop ID in the SCD with the bus stop layer in GIS. Second, we combine the trips of each cardholder to retrieve their full bus travel diary (BTD), which records information on all the trips each cardholder takes each day, as well as their card type. The BTD is the basic data used for further jobs-housing analysis. Following this, the trip count, total bus trip time, total bus trip distance, start point, and end point on each day of the week can be calculated for each cardholder using the generated BTD data. Since the BTD does not include subway rides, cardholders with subway rides were removed from the analysis, according to the "Trip ID" attribute, to avoid identification bias. The jobs-housing relationships are analyzed using the pre-processed BTD data with the approaches below.

We propose two data forms for representing the SCD for each cardholder on each day, trip (TRIP), and position-time-duration (PTD). In TRIP, a trip denotes one record in the SCD, which comprises a cardholder boarding and alighting, namely one bus ride. In the SCD, a trip (TRIP) is stored as its departure location (OP) and time (OT), as well as its arrival stop (DP) and time (DT), as TRIP = {OP, OT, DP, DT}. TRIP is a direct expression of the BTD.

The PTD data form, as an alternative to TRIP, is converted from the TRIP data form and can describe an activity's spatiotemporal characteristics. The generation of PTD assumes that a cardholder does not use travel modes other than bus. For a cardholder, PTD is expressed as PTD = {P, $t$, D}, where P is a bus stop around which the cardholder stays to perform some activity, $t$ is the start time of the activity at location P, and D is the temporal duration at the location P. Compared with TRIP,

PTD better matches the time geography and can identify various types of urban activities. We need to convert TRIP to PTD, and use an example to show how to do so. Assume a cardholder leaves home (bus stop H0) at 7:00 and travels by bus to arrive at their work location (bus stop J0) at 8:00. After working for a full-day, the cardholder leaves their workplace at 17:00 and travels by bus to arrive home at 18:00. The TRIP data form for the two trips is expressed as {H0, 7:00, J0, 8:00} and {J0, 17:00, H0, 18:00}. The converted PTD data form is then expressed as {H0, 18:00 (-1), 13 h} and {J0, 8:00, 9h} and represents two activities, first the home activity and then the work activity. The home activity starts at 18:00 on the previous day and lasts for 13 hours till 7:00. The work activity starts at 8:00 and lasts for 9 hours.

*4.2 Identification of housing and job locations using one-day data*

We use the PTD data form to identify housing and job locations for each cardholder. We hold one-week data, and how to identify housing and job locations using one-day data is the first step of our approach.

To identify housing locations, we suppose the departure bus stop of the first trip (TRIP1) to be the housing location of a cardholder[3]. The housing location is assumed to be within walking distance of the bus stop, which is the spatial resolution of our analytical results. In the 2005 survey, trips with the mode "walking" can be screened to calculate walking duration. The histogram of walking time illustrated in Figure 4 shows that the average walking time is 9.0 min in the 2005 survey for trips with all purposes. Accordingly, the average walking distance is estimated to be 750 m, assuming an average walking speed of 5 km/h (see http://en.wikipedia.org/wiki/Walking). This is the distance a bus rider is likely to walk from their housing or job location to ride a bus, or to reach their destination after riding a bus. Unlike the average size of a TAZ, this distance can be regarded as accurate within a TAZ. In the 2005 survey, 99.5% of first trips started from home, which supports our rule for identifying housing locations. If a cardholder's first trip of the day is on a fixed-fare bus route, we cannot identify the housing location for that cardholder.

---

[3] Notably, there could be a minor bias in identifying the housing location since a cardholder may have taken an illegal taxi to a bus stop prior to their first bus ride.

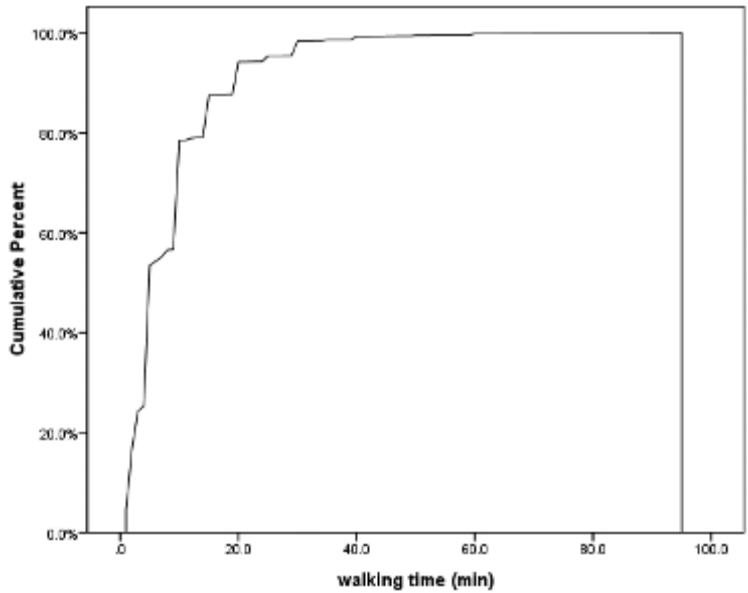

Figure 4 Cumulative probability distribution of trip walking time in the 2005 survey

To identify job locations, we need to identify working trips using bus. Fulltime job locations are identified based on the interval between any two adjacent cardholder trips being long enough for a full time job. This method assumes that the fulltime job activity is the urban activity conducted for the longest time on week days. If a cardholder meets all the conditions below, the $k$th location $P_k$ is regarded as their job location.

*Condition 1*: The card is not a student card.

*Condition 2*: $D_k \geq 360$.

*Condition 3*: $k <> 1$. [4]

That is, for non-student cardholders, if they spend over 360 min (6 h) at any location except the first location, we assume they are working at that location. The benchmark of 6 h is based on the 2005 survey, in which the average working time is 9 h and 19 min (with standard deviation of 1 h and 41 min, see Figure 5 for details) for a sample of 27,550 persons (210 persons went home for a rest at noon, and their data were not counted). Thus 96% of persons sampled work for over 6 h per day. Notably, the process of identifying housing locations is independent from that of identifying job locations.

---

[4] According to the definition of the PTD data form, the first activity of a cardholder is the home-based activity started in the previous day. This rule guarantees the identified job location is not the housing location.

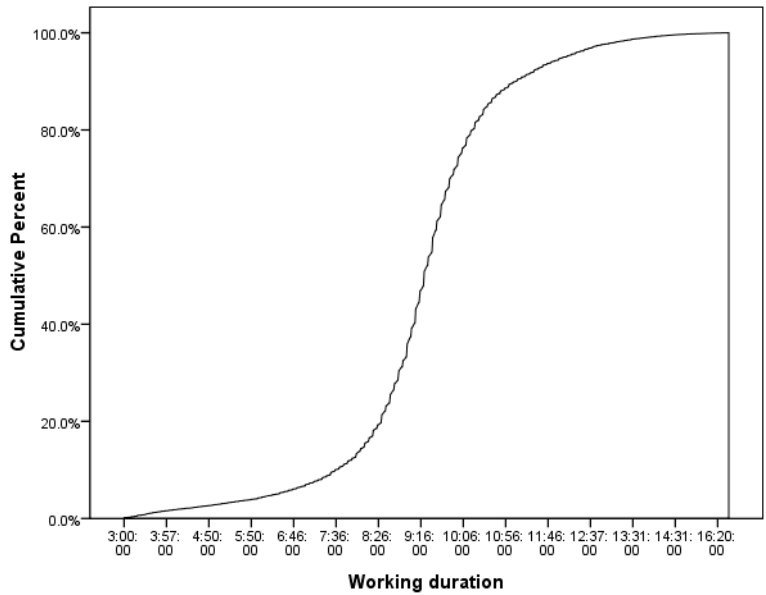

Figure 5 Cumulative probability distribution of working time extracted from the 2005 survey

*4.3 Identification of housing and job locations using one-week data*

Since the periodic patterns of the identified housing or job locations of cardholders vary from each other in the identification results for each day, we need to propose several indicators to combine the one-day results and obtain the final one-week results.

      To determine the final housing locations (that is, the housing locations obtained using the one-week data), we use a rule-based method with a decision tree based on one-day results (see Figure 6). Both the frequency and spatial distribution of identified locations using one-day data are applied in this process. In detail, if a cardholder only visits an identified housing location on one day during the week, we cannot confidently identify this location as the cardholder's housing location. Where multiple identical one-day locations exist, we regard the final location of the cardholder as the housing location. Where multiple different one-day locations exist, they are grouped into clusters according to the distance among them. If the distance between two locations is below a threshold, they are considered to belong to a single cluster. We set this threshold as 500 m, which is about twice the average distance between two adjacent bus stops (231*2=462 m). This threshold is also based on the finding of Zhao et al (2011) that the threshold service distance of bus stops is 500m. If only one cluster exists, the most frequently used location is regarded as the final location. In the case of two locations with the same frequency, the concept of

"residential potential" is introduced to determine a better final location (the concept of "job potential" is also used to determine the final job location). The potential is calculated based on the land use pattern data using Equation (1), in which $p_h^k$ is the housing potential of bus stop $k$, $p_j^k$ is the job potential of bus stop $k$, and the neighborhood of bus stop $k$ is the Thiessen polygon generated from the bus stop layer. Parcels with their centroid inside a neighborhood polygon are defined as the parcels in the neighborhood. The two potential indicators are further rescaled to range from 0 to 1.

$$p_h^k = \frac{\text{the total floor area of housing parcels in the neighborhood of the bus stop } k}{\text{the total floor area of all parcels in the neighborhood of the bus stop } k}$$

$$p_j^k = \frac{\text{the total floor area of job parcels in the neighborhood of the bus stop } k}{\text{the total floor area of all parcels in the neighborhood of the bus stop } k} \quad (1)$$

If there are several clusters and each has only one bus stop, the stop that is the housing location cannot be identified. Further, given only one cluster with maximum locations (the maximum cluster), the most frequent location in this cluster is considered the final location. Otherwise, the most frequent location in the maximum cluster is regarded as the final location; this situation applies when the most frequent location counts are different in maximum clusters. Finally, when the most frequent location counts are the same in the maximum clusters, the maximum location in the maximum cluster with the greatest residential potential is regarded as the final location.

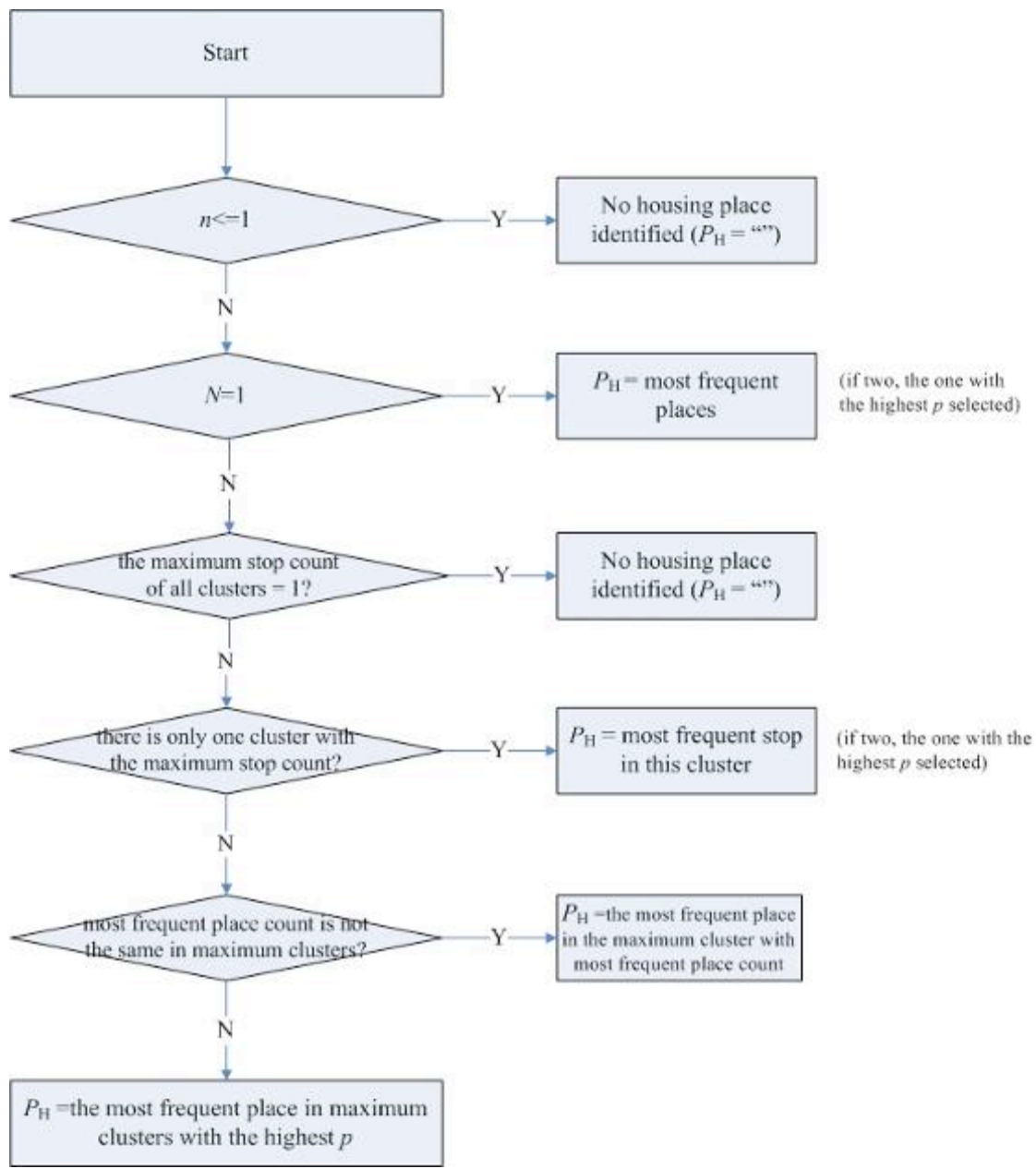

Figure 6 Decision tree diagram for identifying housing locations based on the one-day result

Note that $n$ is the count of identified housing locations for a cardholder during a week, $N$ is the count of clusters of the cardholder, and $P_h$ is a potential housing location for the cardholder. The sequence of identified housing locations is not counted in this decision tree. $p$ is the housing potential of a housing location.

Accordingly, the rule for identifying final job locations is the same as the above rule for housing locations.

*4.4 Identification of commuting trips based on identified housing and job locations*

We use the TRIP data form to identify the commuting trip[5] from a housing location to a job location for cardholders with both an identified housing location and an identified job location using one-week SCD. Commuting distance and time are used as key indicators for measuring commuting patterns. Commuting distance is measured as the Euclidean distance between the housing and job locations. Commuting time is regarded as the time duration between the boarding time at the housing location and the arrival time at the job location. To calculate this requires identifying the commuting trip (corresponding to one or several bus trips/ridings in the TRIP data) from one-week trips where the cardholder meets the following three conditions: (1) The boarding bus stop of the first trip on a given day is the identified housing location. (2) The job location is identified based on trips made during a given day. (3) Both the housing and job locations are identified on the same day (the stops in the same cluster are considered identical in this process). Notably, if a cardholder makes multiple identified commuting trips in a week, commuting time may vary across days, and these different times are averaged to produce a final time for that cardholder. (4) Identified commuting trips with extreme commuting times are removed. The benchmark for identifying extreme commuting times is set to 180 min, which covers 99% of all bus commuting trips in the 2005 survey.

**5 Results**

Two tools were developed for our empirical research in Beijing using the proposed approach. Since the original SCD is stored in the MS SQL Server, data pre-processing and two types of data form building were conducted using a structured query language (SQL) tool in the SQL Server to promote computation efficiency. To effectively use the GIS datasets and spatial analysis and statistics functions, we migrated the processed SCD together with the 2005 survey and GIS layers to a File Geodatabase of ESRI ArcGIS, and developed a Python tool based on ESRI Geoprocessing to identify job and housing locations and commuting trips of cardholders as well as to analyze and visualize commuting patterns using the pre-processed data.

    The results of data analysis are illustrated in Figure 7, which gives a general summary of the information presented in this section.

---

[5] In this paper, both commuting time and commuting distance are calculated based on a one-way trip from the housing location to the job location.

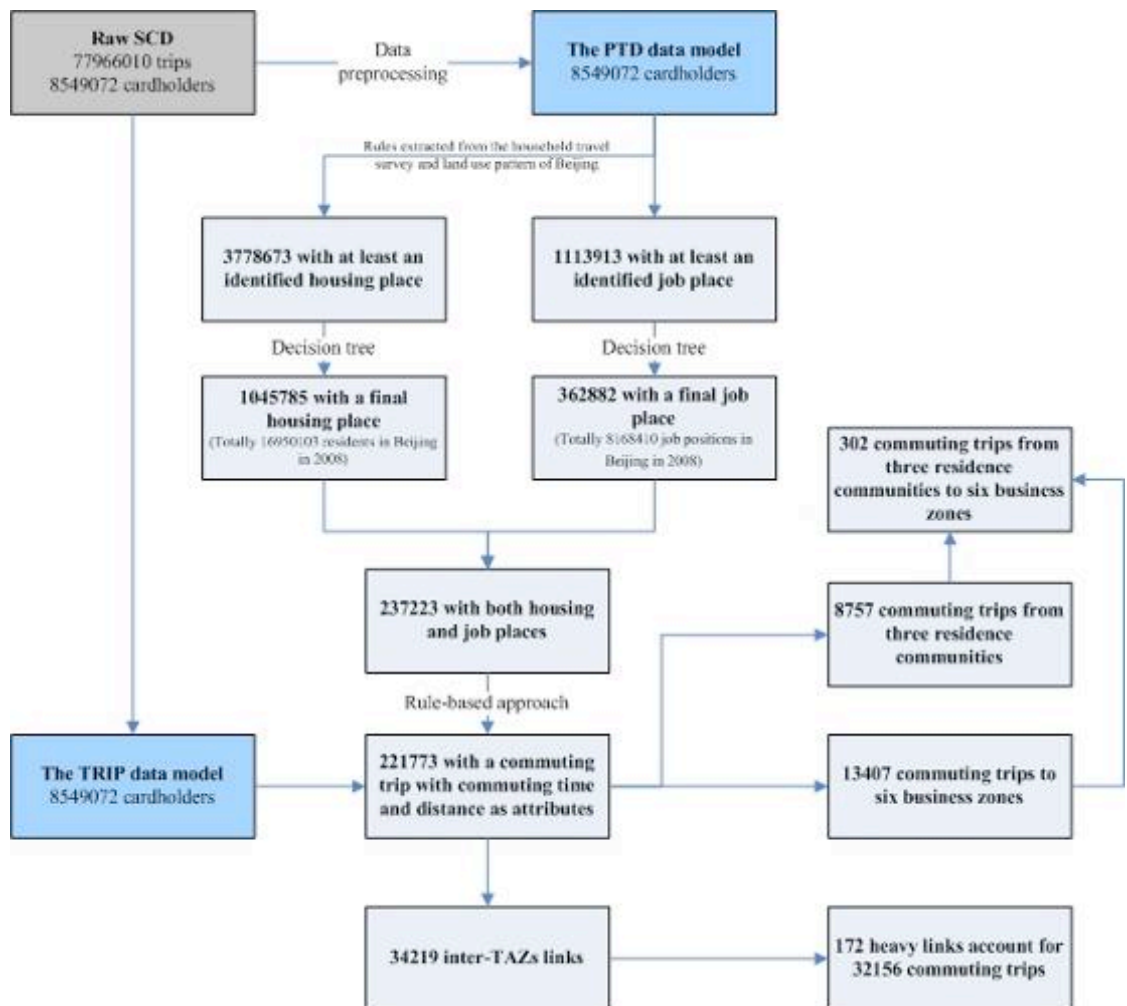

Figure 7 Identification results for the whole paper

## 5.1 Jobs-housing locations identification, aggregation of the bus stop and TAZ scales and comparison with observed data

The jobs-housing locations identification using one-week data is based on the identification results for each day. The periodic pattern in Figure 8 shows that the number of cardholders with identified housing or job locations decreases with the number of days for which identification results existed. Based on the approach in Section 4.3, a final housing or job location can only be obtained for cardholders for whom housing or job locations were identified on two days or more.

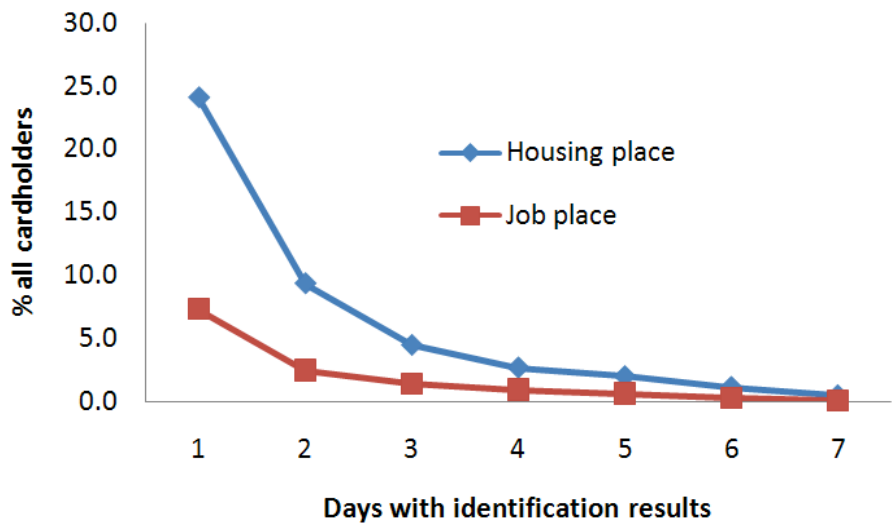

Figure 8 Percentages of cardholders for which identification results exist by day

Using one-week data, housing locations are identified for 1,045,785 cardholders (12.2% of all 8,549,072 cardholders), and job locations are identified for 362,882 cardholders (4.2% of the total). Since the processes of identifying housing and job locations are independent of each other, 237,223 cardholders (2.8% of the total) have both final housing and final job locations.

The identification results are aggregated in the bus stop scale with a total of 3,414 bus stops (39.3% of all 8,691 bus stops) corresponding to housing locations and 3,329 (38.3% of the total) corresponding to job locations. The identification results are further aggregated in the TAZ scale, and 729 among 1,118 TAZs correspond to housing locations. Housing and job kernel density maps (Figure 9) show that both the housing and job densities in the inner area exceed those in the outer area from the perspective of bus landscapes. The housing density map shows a more scattered pattern than the job density map.

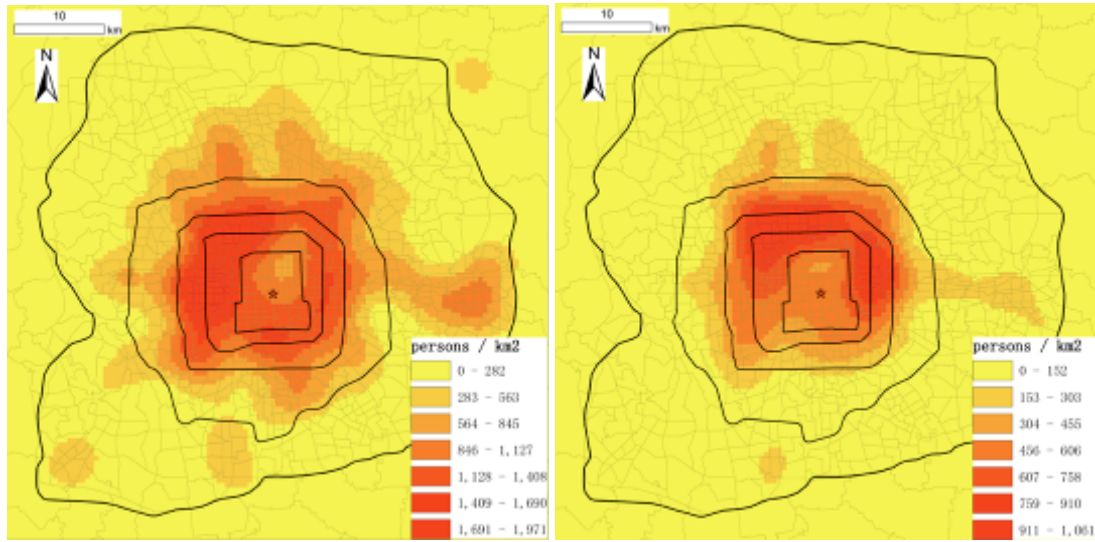

(a)                                   (b)

Figure 9 Identified housing (a) and job (b) kernel density maps in the central BMA

Note: This figure only represents data from bus smart cardholders.

The identification results are compared with the observed housing and job data from 2008 using the TAZ scale. In 2008, the total number of residents in Beijing was 16,950,103, and the number of jobs was 8,168,410 (Beijing Municipal Statistics Bureau and NBS Survey Office in Beijing, 2009). The average identification ratio was 6.2% for housing locations and 4.4% for job locations. Since we do not hold jobs-housing data at the TAZ level in Beijing, we synthesize the 'observed' jobs-housing data for each TAZ using the resident and job numbers for each sub-district from the statistical data in the 2008 yearbook and the floor area of each parcel. Two indicators are used for comparison, the resident ratio (cardholders with identified housing locations in a TAZ divided by observed residents in the TAZ) and job ratio (cardholders with identified job locations in a TAZ divided by observed jobs in the TAZ). The ratio maps are shown in Figure 10ab, in which 729 TAZs have ratio values. There are 29 TAZs with a resident ratio over 100% (Max=16200%), which indicates that the number of bus riders we identified as residing in a TAZ exceeds the number of residents. This phenomenon is considered "abnormal". Regarding the job ratio, 11 TAZs have a job ratio exceeding 100% and the maximum is 10,216%. The median resident ratio among all 729 TAZs is 6.5%, and the median job ratio is 4.3%.

We find that both the resident ratio and job ratio vary significantly among TAZs, which may originate from the spatial heterogeneity of bus share. That is, some TAZs may have greater bus share than other TAZs because of accessibility and the socioeconomic structure of the resident population. We conducted spatial autocorrelation analysis of the two maps. The Moran's I of the resident ratio is -0.0016 and the Z-score is -0.16. The resident ratio pattern thus can be considered randomly distributed. In comparison, the Moran's I of job is 0.0031 and the Z-score is 2.37, which denotes that the job ratio distribution has a clustered pattern and job locations accessed by bus riders tend to agglomerate. Additionally, we conducted Hot Spot Analysis of the two ratio maps to detect specific cluster patterns (see Figure 10cd). Hot spots for the resident ratio are distributed in southern suburban areas, particularly newly-built neighborhoods around the 5$^{th}$ ring road, denoting that a large percentage of residents there commute by bus. For the job ratio, hot spots are distributed mainly in northwest suburban areas around the 5$^{th}$ ring road, namely Shangdi and Huilongguan in Figure 14, denoting that a high percentage of those who work there commute by bus.

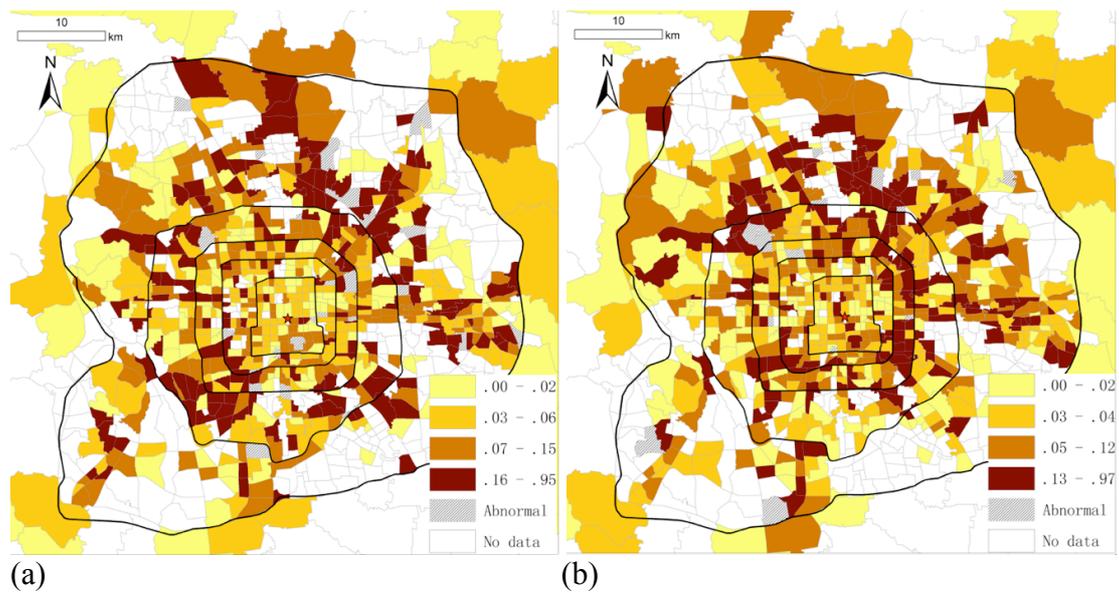

(a)                                      (b)

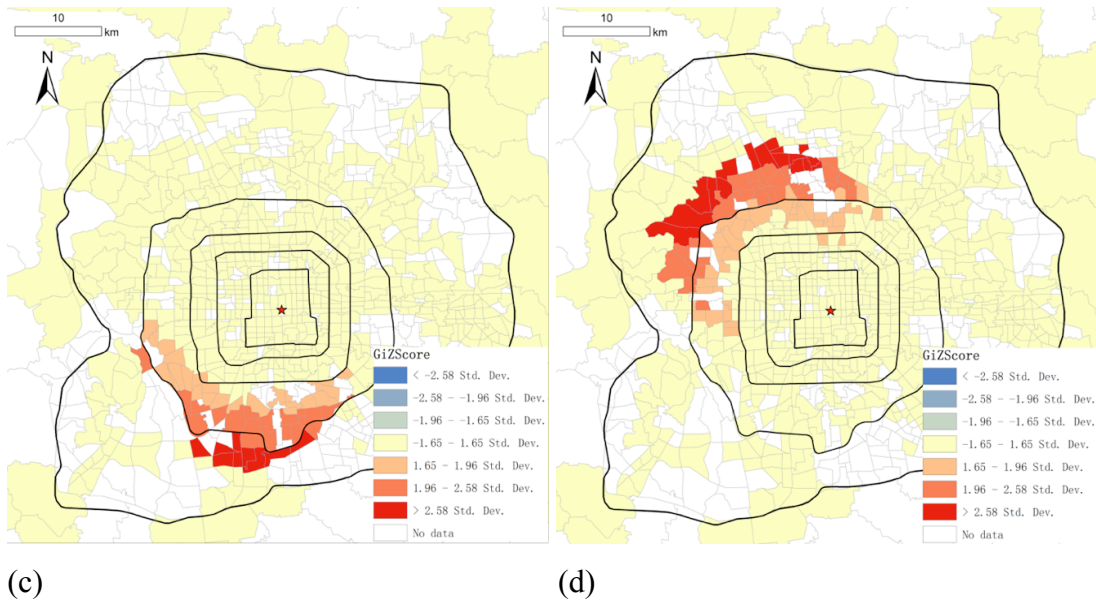

(c)                                        (d)

Figure 10 The bus rider ratio among all residents in terms of resident (a) and job (b) locations as well as hot spot analysis for the resident (c) and job (d) ratios

*5.2 Commuting trip identification and comparison with the 2005 survey and other existing researches*

We identified 221,773 cardholders that made commuting trips using one-week data from 237,223 cardholders for whom both housing and job locations had been identified[6]. The identified commuting trips have average commuting time of 36.0 min and standard deviation of 24.2 min. Meanwhile, the average commuting distance (Euclidean distance) is 8.2 km and the standard deviation is 7.0 km. The average trip time and distance of cardholders are calculated for all TAZs by aggregating the housing locations of identified commuting trips (Figure 11 a and b). For the 729 TAZs with identified commuting trips, the median commuting time is 35.0 min, and the median commuting distance is 7.2 km. TAZs in the central area have lower commuting time and distance than those elsewhere. The circular distribution of commuting distance reflects the mono-centric urban structure of Beijing. To measure the spatial autocorrelation of the calculated average trip time of each TAZ, we calculated the global Moran's I indicator. Moran's I is 0.024 with Z Score of 11.57, indicating less than a 1% likelihood that this clustered pattern could occur randomly. For commuting distance, Moran's I is 0.067 with Z score of 31.08. Therefore, TAZs are significantly clustered in terms of both average trip time and average distance.

---

[6] Cardholders that undertake commuting trips slightly less than those with both housing and job locations, which reflects the fact that housing and job locations must be identified on the same day for us to identify a commuting trip for the cardholder, as elaborated in Section 4.4.

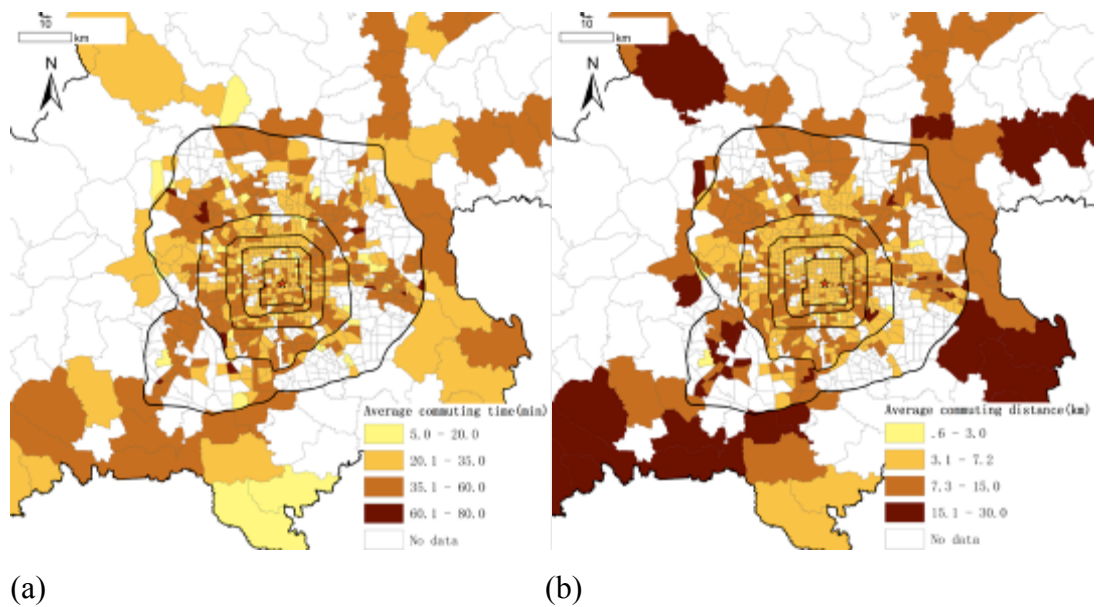

(a)　　　　　　　　　　　　　　(b)

Figure 11 Average commuting time (a) and distance (b) for each TAZ in Beijing

We compared our identification results with the 2005 survey to validate our approach (see Table 4). The 2005 survey contained 6,651 persons who commute by bus, or 3% of the number we identified. The 2005 survey included three levels of validation, as follows: (1) The average commuting time was 40.5 min (Std.D=23.1), and the average commuting distance (both going to work and returning home) was 8.4 km (Std.D=8.3 km) in the 2005 survey. The $t$-test between the 2005 survey and our results reveals a significant difference ($t$=15.7). This may result from the sample size and strategy of the survey, as well as from potential bias of the data-mining algorithm. (2) We further compare the cumulative distribution function (CDF) of commuting trips in the 2005 survey with our results (Figure 12). The CDFs of commuting time generally overlap between our results and the 2005 survey, although the CDF is not smooth for the latter because commuting time was recorded by commuter memory and discretized into various categories. For commuting distance the CDFs overlap almost perfectly.

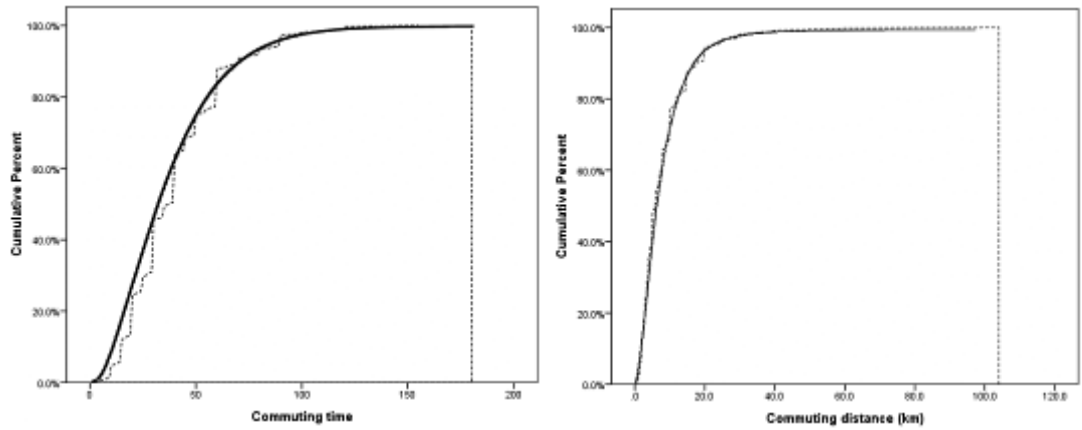

(a)                                             (b)

Figure 12 Comparison of CDFs of commuting time (a) and distance (b) between commuting trips in the 2005 survey (dashed lines) and our identified results (solid lines)

(3) Since the commuting trip count in the 2005 survey is not large enough to conduct the comparison in the TAZ scale, the comparison is instead conducted in the district scale. The BMA includes 18 districts, as shown in Figure 1d, including eight in the central area, five in the near suburban area and five in the remote suburban area. The comparison shows that our results coincide closely with those of the survey, especially in the central area where most trips are generated. This is also supported by Wilcoxon signed-rank tests for both commuting time ($p$=0.286) and distance ($p$=0.267) of 18 districts. A difference between our results and those of the survey (e.g. the commuting time ratio of 1.54 for the Huairou district), may lie in the limited sample size in this district in the 2005 survey. Furthermore, OD pair frequencies of identified commuting trips at the TAZ level suggest a typical power-law distribution, which further proved the applicability of our approach. To summarize, validations prove the applicability of our proposed approaches for identifying housing and job locations as well as commuting trips.

Table 3 District-scale comparison for commuting time ($t$) and distance ($d$) of commuting trips between our identification results and the 2005 survey

| District | | | Our results | | | The 2005 survey | | | Our results/survey results | |
|---|---|---|---|---|---|---|---|---|---|---|
| | | | Count | $t$ (min) | $d$ (km) | Count | $t$ (min) | $d$ (km) | $t$ ratio | $d$ ratio |
| Central area | | Dongcheng | 4,179 | 35.1 | 6.5 | 317 | 37.7 | 5.8 | ***0.93*** | ***1.12*** |
| | | Xicheng | 9,145 | 33.7 | 7.1 | 467 | 35.2 | 6.3 | ***0.96*** | ***1.13*** |

| District | | Our results | | | The 2005 survey | | | Our results/survey results | |
|---|---|---|---|---|---|---|---|---|---|
| | | Count | t (min) | d (km) | Count | t (min) | d (km) | t ratio | d ratio |
| | **Chongwen** | 3,762 | 39.8 | 7.6 | 276 | 37.6 | 5.8 | *1.06* | *1.31* |
| | **Xuanwu** | 4,377 | 36.6 | 8.2 | 432 | 40.3 | 6.9 | *0.91* | *1.19* |
| | **Chaoyang** | 66,918 | 37.2 | 7.5 | 2031 | 42.7 | 8.7 | *0.87* | *0.87* |
| | **Haidian** | 48,888 | 35.7 | 7.3 | 1277 | 39.8 | 8.0 | *0.90* | *0.92* |
| | **Fengtai** | 32,170 | 38.6 | 9.0 | 678 | 46.6 | 9.9 | *0.83* | *0.91* |
| | **Shijingshan** | 4,561 | 34.3 | 7.6 | 313 | 30.3 | 6.2 | *1.13* | *1.21* |
| Near suburbs | **Changping** | 13,035 | 36.5 | 8.8 | 202 | 47.4 | 11.1 | *0.77* | *0.79* |
| | **Tongzhou** | 10,400 | 38.4 | 10.1 | 181 | 40.9 | 12.8 | *0.94* | *0.79* |
| | **Daxing** | 9,455 | 38.9 | 9.1 | 94 | 40.1 | 10.1 | *0.97* | *0.91* |
| | **Fangshan** | 3,057 | 47.4 | 15.7 | 157 | 31.7 | 11.5 | *1.49* | *1.37* |
| | **Mentougou** | 1,196 | 31.1 | 9.9 | 113 | 36.7 | 9.1 | *0.85* | *1.08* |
| Remote suburbs | **Huairou** | 299 | 44.3 | 12.5 | 8 | 28.8 | 11.6 | *1.54* | *1.08* |
| | **Miyun** | 149 | 43.7 | 13.1 | 7 | 34.6 | 16.1 | *1.26* | *0.82* |
| | **Pinggu** | 730 | 43.8 | 15.7 | 8 | 42.5 | 23.8 | *1.03* | *0.66* |
| | **Shunyi** | 5,497 | 34.3 | 10.0 | 80 | 39.5 | 14.1 | *0.87* | *0.71* |
| | **Yanqing** | 254 | 36.8 | 12.1 | 10 | 56.0 | 41.9 | *0.66* | *0.29* |

We also compare our results with the existing research, as shown in Table 4. Besides the 2005 survey, the other three scholarly surveys did not record commuting distance. Commuting times in the three surveys significantly exceed our results. This phenomenon may reflect the improvement of the Beijing public transportation system from 2001 to 2008. However, another possible explanation is that our commuting times exclude walking or cycling time from the housing location to the bus stop and from the bus stop to the job location.

Table 4 Commuting time and distance data for Beijing based on existing studies

| Name of Study | Travel modes and year of research | Sample size | Average commuting time (min) | Average commuting distance (km) |
|---|---|---|---|---|
| **Present study** | **Bus, 2008** | **221,773** | **36.0 (24.2)** | **8.2 (7.0)** |
| The 2005 survey | Bus, 2005 | 6,651 | 40.5 (23.1) | 8.4 (8.3) |

| Name of Study | Travel modes and year of research | Sample size | Average commuting time (min) | Average commuting distance (km) |
|---|---|---|---|---|
| Liu and Wang, 2011 | Bus, 2007 | 307 | 46.3 (N/A) | N/A |
| Wang and Chai, 2009 | Bus, 2001 | 227 | 55.1 (30.4) | N/A |
| Zhao et al, 2011 | Bus and metro, 2001 | 220 | 52.4 (26.6) | N/A |

Note that the numbers in brackets are the standard deviation of the average commuting time and distance. Bus samples in studies other than the present one are extracted from the survey of all travel modes.

*5.3 Commuting trips visualization for the whole region and typical zones*

Mapping identified commuting trips is an effective method of understanding commuting patterns in Beijing. Each commuting trip is visualized as a line that links the departure (housing) and arrival (job) bus stops, and has associated commuting time, commuting distance and card ID as GIS attributes. All commuting trips in the stop scale are available in our GIS database. To identify the dominant commuting pattern in the BMA, we further aggregate commuting trips into the TAZ scale and obtain the trip counts between different pairs of TAZs (we term such trips links in this paper). The inter-TAZ commuting pattern contains 34,219 links.

We calculated the OD direction for each inter-TAZ link. The direction indictor, increasing counter-clockwise, ranges from 0 to 360 degrees, with 0 denoting an easterly direction. The average direction for each TAZ was calculated by aggregating all links departing from a TAZ. Figure 13a shows the results and reveals a centripetal pattern, indicating that commuting to the center (Tian'anmen) is the dominant phenomenon. We further evaluated the pattern using global spatial autocorrelation and showed a cluster pattern (Moran's I=0.11 and Z-score=46.65). Local Indicators of Spatial Association (LISA) results shown in Figure 13b illustrate that HH and LL clusters are common in the central area. The LL cluster, consisting of TAZs with a direction from 0 to 90 degrees, lies to the south west of the center. The HH cluster is to the north of the center. The two clusters further confirm the centripetal commuting pattern.

(c) We divided the direction into eight categories, separated at 45 degree intervals in the range 0 to 360 degrees. The frequencies of the various categories were then calculated for each TAZ. According to the composition of the category frequencies of each TAZ, we clustered all 730 TAZs into four clusters using the "K-Means Cluster Analysis" approach in SPSS (see Figure 13c). The four clusters include 214, 165, 224, and 126 TAZs, respectively, and highlight various dominant commuting directions. The dominant direction of Cluster 1 is 180 to 315 degrees, Cluster 2 has two equally dominant directions in 0 to 45 and 315 to 360 degrees, the dominant direction for Cluster 3 is 90 to 180 degrees, and that for Cluster 4 is 45 to 90 degrees.

We use the head/tail division rule proposed by Jiang and Liu (2011) to divide all links into six levels based on their trip counts (see Figure 13d, which shows links between the centroids of TAZs). In levels 4 to 6 with heavy commuting traffic, 175 links (0.5% of all links) account for 32,156 commuting trips (14.8% of all trips). These heavy links are mainly located within the 6$^{th}$ ring road and express public transportation services are recommended to service the few heavy links that cross this ring road. Figure 13d illustrates the dominant commuting patterns in the BMA.

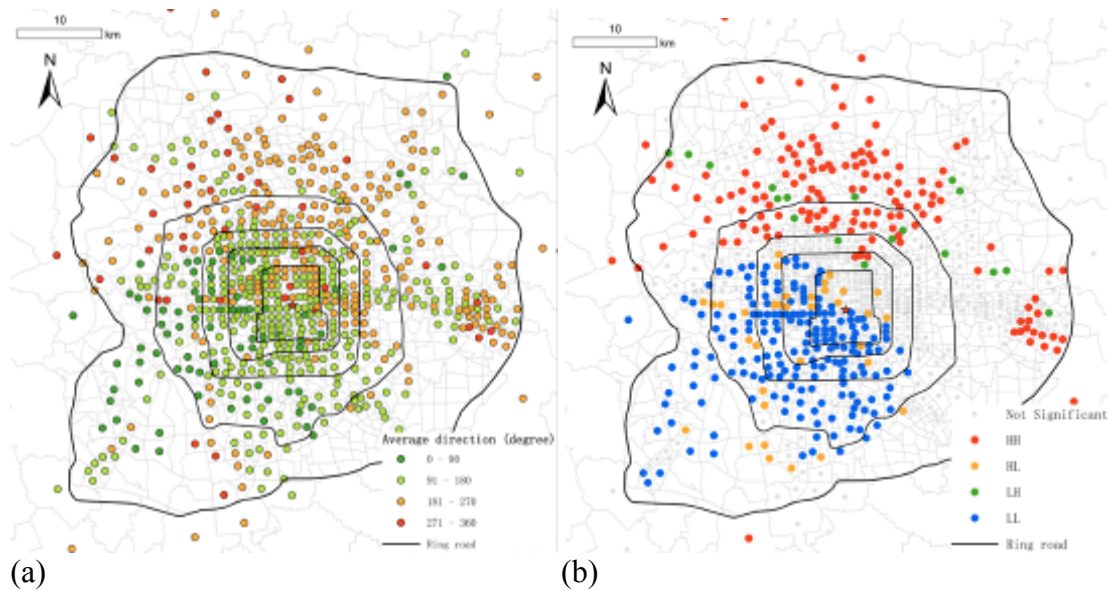

(a)  (b)

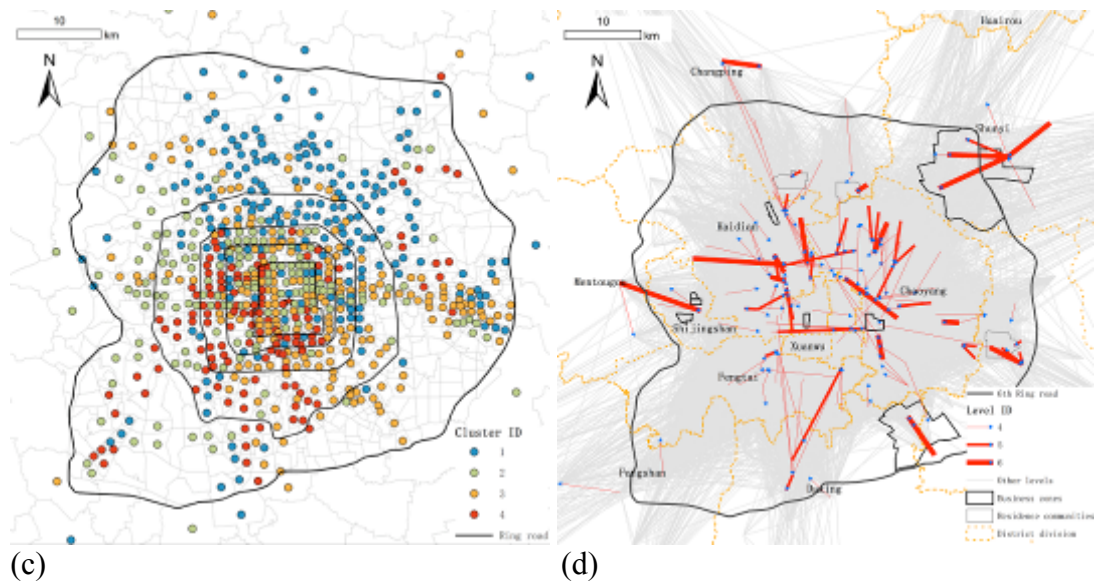

(c)                                                      (d)

Figure 13 Commuting directions in central Beijing: (a) the dominant commuting direction for each TAZ; (b) LISA results; (c) cluster analysis results; (d) links in the TAZ scale illustrating commuting patterns

Note: Arrows in map (d) denote the commuting direction from housing location to job location.

The congestion Beijing suffers from 'tidal traffic' caused by overly large residential communities and overly agglomerated business zones is widely discussed in the Chinese media. We extract commuting trips originating from housing locations within three major residential communities, Huilongguan, Tiantongyuan and Tongzhou (see Figure 14a). The former two zones are the biggest communities in Northern Beijing and were built in the 1990's, while the Tongzhou area in Eastern Beijing contains several newly-built residential communities. Similarly, we extract commuting trips with job locations in six dominant business zones (Figure 14bc), the CBD, Shangdi (an IT park), Yizhuang (the biggest industrial zone), Tianzhu (the airport base), Shijingshan (an business zone in western Beijing) and Jinrongjie (a financial, banking and insurance zone).

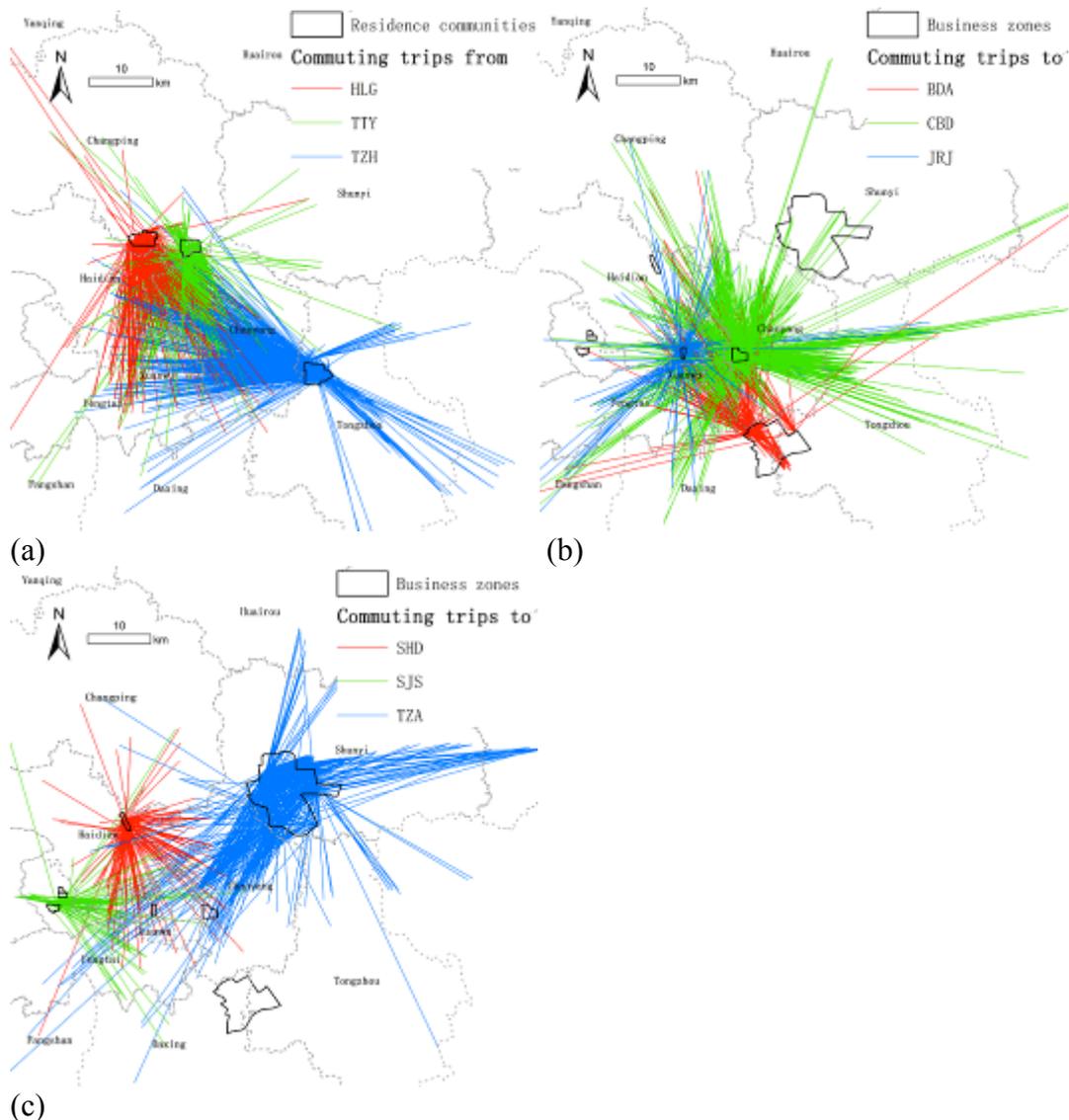

(a)　　　　　　　　　　　　　(b)

(c)

Figure 14 Typical commuting trips a) from three main residential communities; b and c) to six main business zones

Note: HLG = Huilongguan community, TTY=Tiantongyuan community, TZH=Tongzhou community, CBD=Central Business District, SHD=Shangdi Industrial Park, BDA=Beijing Development Area at Yizhuang, TZA=Tianzhu Airport Park, JRJ=Jinrongjie (financial zone), SJS=Shijingshan Park

Commuting trips from each community or to each zone are further aggregated in terms of commuting time and distance (Table 5). Residents of TTY tend to commute shorter distances than those of TZH, and few of either group work in southern Beijing. Some residents of TZH work in new cities on the outskirts of Beijing, and a few work in western Beijing. Regarding commuting trips to business zones, the CBD attracts workers from more geographically extensive locations, resulting in the highest average commuting time of all business zones. Workers in BDA have the least commuting time and distance, which may result from its local job center status. Surprisingly, only 302 commuting trips (0.14% of the total number identified) are from the three major residence communities to the six major business zones in Beijing. Figure 13d supports this unexpected finding. One reason for this phenomenon may be that most commuters moving between these zones drive cars rather than travel by bus[7].

Table 5 Commuting time and distance for various residential communities and business zones in Beijing

| Zone name | Commuting time (min) | Commuting distance (km) | % of all identified commuting trips |
|---|---|---|---|
| Trips from residential communities | | | 3.9 |
| TZH | 45.1 | 10.0 | 1.4 |
| HLG | 39.4 | 7.0 | 1.0 |
| TTY | 36.2 | 6.1 | 1.5 |
| Trips to business zones | | | 6.0 |
| CBD | 41.4 | 9.4 | 2.7 |
| SHD | 40.4 | 6.7 | 0.3 |
| JRJ | 34.9 | 7.1 | 0.5 |
| TZA | 31.6 | 10.0 | 1.3 |
| SJS | 28.4 | 6.9 | 0.3 |
| BDA | 26.6 | 6.4 | 0.8 |

## 6 Discussion

### 6.1 Our contributions

This paper makes three main contributions. **First**, we investigate urban dynamics based on ubiquitous LBS data using rules generated from conventional surveys and GIS layers, which is a promising approach to analyzing LBS data (Batty, 2012). Comparison with the 2005 survey has validated this approach on three levels, showing a significant and sound result. This can combine the spatiotemporal dimensions of rich LBS data with the social dimensions of conventional surveys to better understand urban dynamics. Additionally, the identification results obtained from the SCD can provide useful information during the multi-year gaps that separate surveys due to their demanding nature in terms of human and financial resources. **Second**, a whole week of SCD analysis, tracking individual cardholder bus trips, is applied to analyze commuting patterns in Beijing, thus making our identification results more solid than those based on one-day data. We also proposed a decision tree for determining the final one-week results using the periodic information and spatial distribution of the one-day results. **Third**, we retrieved explicit spatial commuting patterns for Beijing based on more accurate information than conventional questionnaires or household travel surveys. To our knowledge, our commuting pattern analysis of Beijing involves a larger sample size and more precise spatial and temporal information than any previous studies, although it is limited to bus riders and represents only a small proportion of the residents of Beijing. This paper highlights policy implications for urban planning and transportation planning based on identification of commuting patterns. In sum, our test use of SCD is promising for analyzing urban dynamics, especially commuting patterns, offers an alternative solution for analyzing commuting issues in a mega region, and can offset some of the weaknesses of conventional surveys.

---

[7] To alleviate traffic congestion, the Beijing government started operating commuting shuttle buses (both morning and evening services) linking HLG and TTY with JRJ on April 26, 2011 (http://news.sina.com.cn/c/2011-04-26/013522356172.shtml), and linking TZH with the CBD on June 14, 2011 (http://news.dichan.sina.com.cn/bj/2011/06/14/333255.html). The government is now planning to operate more shuttle bus lines linking the major residential communities and major business zones mentioned in this paper. The heavy links shown in Figure 13d support such policies.

With the booming availability of information technology, large volumes of individual commuting data are increasingly ubiquitous, thus making individual travel diaries available for further data mining and decision support. This study represents a typical application of this sort of data. The SCD we obtained for a one week period in Beijing totals 21 GB. Data pre-processing and data form building in the SQL Server takes 8 hours, and jobs-housing identification (72 min) and commuting trips identification (113 min) using our Python tool. This research was conducted using a workstation with a CPU of 3.0 GHz * 2 and memory of 4 GB.

*6.2 Limitations of the smart card data*

While we appreciate the rich information provided by the smart card data used in this paper, applying SCD to investigate urban systems suffers several limitations. First, SCD is limited to bus riders and trips using other modes are excluded. Future studies should attempt to combine SCD with all-modal travel surveys. Second, the spatiotemporal information of SCD generated by fixed-fare bus routes in this paper is incomplete. Most short fixed-fare routes are distributed in the central area, while longer distance-fare routes are distributed across the central and outskirt areas. Thus part of the spatial and temporal information of bus rides in the central area is lost, which means identification results, and hence policy implications, are more accurate and complete in outskirt areas. Third, bus trips paid for by cash and cases of card sharing are not counted, although in Beijing they comprise only a small ratio of total trips. Fourth, the anonymity of the smart cards in this study prevents the inclusion of socio-demographic information in SCD, making it hard to conduct behavioral study at the cardholder level. The above limitations of SCD may be addressed in future studies involving more comprehensive SCD from cards that store more information.

*6.3 Future works*

In the near future, we plan to expand on this paper as follows. **First**, smart cards are also widely used in the metro system in Beijing, and the data format is the same as the SCD used in this paper. The metro's share of total journeys relative to all modes of household transportation in Beijing has recently increased, from 8.0% in 2008 to 10.0% in 2009, stimulated by rapid construction of subway lines. In future we will try to acquire SCD for metro lines to get more complete passenger travel data and identify more realistic commuting patterns. **Second**, PTD can be used to study other urban activities (e.g. shopping, hospital and recreation) besides the home and work activities considered in this paper. The 2005 survey and trip purpose information can be applied to retrieve rules for identifying various urban activities through a similar approach to that used in this paper.

**7 Concluding remarks**

This paper demonstrated a preliminary step toward using SCD for urban jobs-housing relationships analysis, including evaluating spatiotemporal dynamics of bus commuting system, identifying jobs-housing locations and commuting trips, and analyzing commuting patterns in terms of time and distance.

**First**, we proposed two data forms, the original TRIP and location-time-duration (PTD), for processing and mining the raw SCD. The PTD data form is for identifying housing and job locations, and the TRIP data form is for identifying commuting trips.

**Second**, we proposed an algorithm for identifying housing and job locations using one-day data based on rules extracted from the 2005 survey and land use patterns in Beijing. We then used a decision tree to combine the one-day results to retrieve one-week results, thus providing more accurate identification results. Finally, housing locations were identified for 1,045,785 cardholders, job locations for 362,882 cardholders, and both housing and job locations for 237,223 cardholders.

**Third**, commuting trips were identified and further mapped based on identified housing and job locations. In total there were 221,773 cardholders with identified commuting trips. Commuting time and distance were averaged and aggregated in the TAZ scale to present the overall commuting patterns in Beijing. We analyzed commuting trips from three typical residential communities and six typical business zones to illustrate the 'tidal traffic' phenomenon in Beijing, an analysis that represents the first explicitly spatial test of commuting patterns in Beijing. We also aggregated commuting trips on the TAZ scale and generated links between pairs of TAZs by recording the commuting trip count. Dominant links were identified to demonstrate the mainstream commuting patterns. Both forms of analysis can obtain information useful to urban and transportation planners and decision makers.

**Fourth**, we validated our identified commuting trips on three levels (the average and standard deviation values, cumulative distribution function and spatial distribution in the district scale) by comparing them with those in the 2005 survey in terms of commuting time and distance. The sound validation result proves the applicability of our approach.

The findings of this study demonstrate the feasibility of spatiotemporal analysis of urban structure using smart card data as an alternative to conventional household travel surveys. This paper also tests novel methods of identifying interesting information from massive geo-tag datasets using rules retrieved from conventional questionnaires or surveys (e.g. the Beijing household travel survey) and urban GIS datasets (e.g. land use pattern, bus stops, and TAZs). Future research can further develop and highlight such novel methods for using ubiquitous geo-tagged data.

**Acknowledgments**: We would like to acknowledge the financial support of the National Natural Science Foundation of China (No.51078213). Zhenjiang Shen at Kanazawa University is also appreciated for his valuable comments on an early draft of this paper. The online version of this manuscript includes a sample of bus smart card data for Beijing we released.